# Reaching and violating thermodynamic uncertainty bounds in information engines

Govind Paneru[1], Sandipan Dutta[1], Tsvi Tlusty[1,2†], and Hyuk Kyu Pak[1,2*]

[1]*Center for Soft and Living Matter, Institute for Basic Science (IBS), Ulsan 44919, South Korea*
[2]*Department of Physics, Ulsan National Institute of Science and Technology, Ulsan 44919, South Korea*

(Dated: August 31, 2020)

Thermodynamic uncertainty relations (TURs) set fundamental bounds on the fluctuation and dissipation of stochastic systems. Here, we examine these bounds, in experiment and theory, by exploring the entire phase space of a cyclic information engine operating in a non-equilibrium steady state. Close to its maximal efficiency, we find that the engine violates the original TUR. This first experimental demonstration of TUR violation agrees with recently proposed softer bounds: The engine satisfies two generalized TUR bounds derived from the detailed fluctuation theorem with feedback control and another bound linking fluctuation and dissipation to mutual information and Renyi divergence. We examine how the interplay of work fluctuation and dissipation shapes the information conversion efficiency of the engine, and find that dissipation is minimal at a finite noise level, where the original TUR is violated.

*Introduction.—* The progress of stochastic thermodynamics in the last decades has borne fruit in the form of new universal laws, such as fluctuation theorems [1-5], which apply to various far-from-equilibrium systems, artificial and living [6,7], and were experimentally tested in several cases [8-12]. A seminal result in this field is the recently discovered thermodynamic uncertainty relation (TUR) [13-17]. In analogy to Heisenberg's uncertainty principle, the TUR sets a fundamental lower bound on the interplay between fluctuation and dissipation in stochastic systems, which originates from the first principles, the inherent thermal fluctuations. The seminal TUR was confirmed theoretically in several stochastic systems [17], and was found to affect the performance of molecular engines [18] and biological synthesis circuits [19].

Recent studies suggest that the TUR bound is satisfied by specific classes of stochastic processes driven by non-equilibrium and time-independent forces that do not change sign under time reversal [17]. And several attempts have been made to identify alternative lower bounds in terms of generalized TURs (GTURs) [20-24]. The TUR has been experimentally tested in non-feedback systems, such as heat engines [25]. But so far, there was no experimental study exploring the validity of the TUR and other bounds in information engines [4,12,26-30], which use measurement and feedback control to extract work from the information on the microstate of a stochastic system.

The TUR has special relevance to stochastic engines, and in particular to information engines: A major question in this field is how to optimize the engines such that the fluctuations in their power and the energy they dissipate are minimal. A direct outcome of the TUR is that these two performance measures, fluctuation and dissipation, cannot be minimized independently, as they are constrained by a general tradeoff. Another lower bound [31], links the fluctuation-dissipation tradeoff of information engines to the mutual information and the Renyi divergence, an information-theoretic distance between the equilibrium and non-equilibrium distributions (hence denoted as IDR, Information Distance Relation). Like the TUR, the IDR is also not well explored experimentally, probably due to the challenge of measuring the fluctuations of mutual information and extracted work.

All these motivated us to test the validity of these universal lower bounds by examining the fluctuation-dissipation tradeoff in information engines. To this end, we constructed a cyclic information engine made of an optically trapped colloidal particle. We explored the entire phase space of the engine, deep into the far-from-equilibrium regime. Our apparatus can measure the noise – and thereby the fluctuations of the thermodynamic variables – very accurately, and the measurements agreed well with a simple theory. We found that in certain regions of the engine's phase space, the system violates the original TUR bound. Yet, the softer version of the generalized bounds derived from the generalized detailed fluctuation theorems, the GTURs [22,23,32,33], still holds. Note that the original TUR is found to be valid in non-equilibrium steady-state under constant driving. In contrast, our system reaches a periodic steady-state with a time-dependent feedback-controlled driving. Indeed, we found that with an appropriate backward protocol, the original TUR is satisfied in a wider phase space and the GTURs are satisfied for the entire phase space of the engine. We also discuss how an optimal protocol achieves the Renyi-information lower bound (IDR) [31]. Lastly, we found that dissipation is minimal near the noise level where the efficiency peaks and the TUR is violated, suggesting that the fluctuation-dissipation tradeoff is the underlying reason for the maximal efficiency at finite noise. Overall, our experiment provides the first test of the original TUR and the other tradeoff relations for feedback systems.

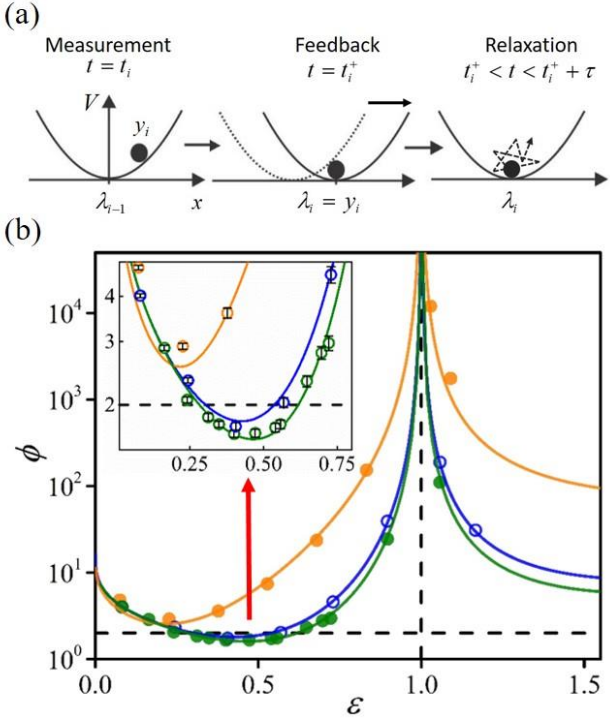

FIG. 1. (a) The engine cycle for symmetric feedback control protocol. At the beginning of the $i$-th cycle, the particle is located at $x$ w.r.t. the trap center $\lambda_{i-1}$. The demon measures the $x$ as $y = x + \text{error}$, with a normally-distributed error of variance $N$. Basing on the measured $y_i$, the demon performs the feedback control step by instantaneously shifting the trap center to $\lambda_i = y_i$. The particle then relaxes for a time $\tau$ in the shifted potential until the next cycle begins. (b) Test of thermodynamic uncertainty relation. The measured fluctuation-dissipation product
$\phi = [\text{Var}(\beta W)/\langle \beta W \rangle^2][\langle \beta W \rangle + \langle I \rangle]$ as a function of error level $\varepsilon = \sqrt{N/S}$ for $\tau = 20$ (olive circles), 3 (blue), and 0.5 (orange) ms, for symmetric feedback. The solid curves are the theoretical values of $\phi$ (Eqs. (1)-(3)). The dashed horizontal line is the TUR lower bound, $\phi \geq 2$. Inset: An expanded view of the region where $\phi$ falls below the TUR bound (olive and blue). The error-bars of $\phi$ (black whiskers) are about the same size of the symbols.

*The mutual information engine.*— In the following, we briefly revisit the information engine and its basic energetics, which we use to test the uncertainty bounds. The engine consists of a colloidal particle immersed in a bath of temperature $k_B T = \beta^{-1}$ and diffusing in the harmonic potential $V(x,t) = (k/2)[x - \lambda(t)]^2$ generated by an optical trap (See [34], for experimental methods). Here, $x$ is the particle position at time $t$, $k$ is the trap stiffness, and $\lambda(t)$ is the center of the trap. Each engine cycle of period $\tau$ includes (i) measurement of the particle position, (ii) shift of the potential center, and (iii) relaxation. We employ two types of feedback control protocols: symmetric and asymmetric. Figure 1(a) shows the schematic of the $i$-th engine cycle under *symmetric* feedback control [30]. Here, the demon measures the *true* particle position $x_i$ with respect to the potential center $\lambda_{i-1}$. But due to Gaussian noise of variance $N$, the demon receives an inaccurate measurement *outcome* $y_i$. The trap center is then shifted instantaneously (that is very fast) to $y_i$, and the particle relaxes for the duration $\tau$ before the next cycle begins. In the subsequent $(i+1)$-th cycle, the particle position is measured with respect to the shifted potential center $\lambda_i$ (the origin is reset) and the same protocol is repeated. Since the origin is reset, the process does not depend on all previous measurements. In the *asymmetric* feedback control protocol, the trap center is shifted to $y_i$ only if $y_i \geq \lambda_{i-1}$, and otherwise remains at $\lambda_{i-1}$ until the next cycle begins.

The dynamics of the particle during the relaxation is described by the overdamped Langevin equation [12,35]. Without feedback, the particle position follows the Gaussian equilibrium distribution of variance $S$ from which we calibrate the trap stiffness as $k = (\beta S)^{-1}$. The characteristic time it takes for the particle to relax towards equilibrium is $\tau_R = \gamma/k \approx 3.5$ ms, where $\gamma$ is the Stokes friction coefficient. After repeating the feedback cycle many times, the system approaches a steady state. For the *symmetric* feedback scheme, the steady state probability distributions of the particle position $p(x)$ and measurement outcome $p(y)$ are also Gaussian [30]. The work performed on the particle, when the potential is shifted, is $\beta W \equiv \beta \Delta V = (1/2)\beta k[(x-y)^2 - x^2]$. Therefore the averge work performed on the particle per cycle in steady-state $\langle \beta W \rangle$ and its standrad deviation std($\beta W$) are

$$\langle \beta W \rangle = \frac{N - S^*}{2S} \quad \text{and} \quad \text{std}(\beta W) = \sqrt{\frac{N^2 + S^{*2}}{2S^2}}, \quad (1)$$

where $S^*(\tau) = S + (N - S)\exp(-2\tau/\tau_R)$ is the variance of $p(x)$. During the relaxation, the steady state average heat supplied to the system $\langle \beta Q \rangle$ is minus the average work performed, $\langle \beta Q \rangle = -\langle \beta W \rangle$. Similarly, the steady state average gain of mutual information per cycle $\langle I \rangle$, between the true particle position $x$ and the measurement outcome $y$, and its standard deviation std($I$) are

$$\langle I \rangle = \frac{1}{2}\ln\left(1 + \frac{S^*}{N}\right) \quad \text{and} \quad \text{std}(I) = \sqrt{\frac{S^*}{S^* + N}}. \quad (2)$$

*Testing the thermodynamic uncertainty relation (TUR).*— In a Markovian and overdamped system driven into nonequilibrium steady state by time-independent forces, the TUR bound on a current $X(t)$ is constrained by $[\text{Var}(X(t))/\langle X(t) \rangle^2]\langle \sigma \rangle \geq 2$, where $\langle \sigma \rangle$ is the average total entropy production [13,14,25]. According to the generalized second law of thermodynamics, the entropy production per cycle in a system with measurement and feedback control includes three contributions, $\langle \sigma \rangle = \langle \Delta S_{\text{sys}} \rangle + \langle \Delta S_m \rangle + \langle \Delta I \rangle$, where $\Delta S_{\text{sys}}$ is the system entropy change, $\Delta S_m$ is the bath entropy change, and $\Delta I$ is the net information gain per cycle [36]. For the current protocols, $\langle \Delta S_{\text{sys}} \rangle = 0$, $\langle \Delta S_m \rangle = -\langle \beta Q \rangle = \langle \beta W \rangle$, and $\langle \Delta I \rangle = \langle I \rangle$. The TUR for the average power (i.e. work current) per cycle, $P \equiv \langle \beta W \rangle / \tau$ then becomes

$$\phi \equiv \frac{\text{Var}(\beta W)}{\langle \beta W \rangle^2}(\langle \beta W \rangle + \langle I \rangle) \geq 2. \quad (3)$$

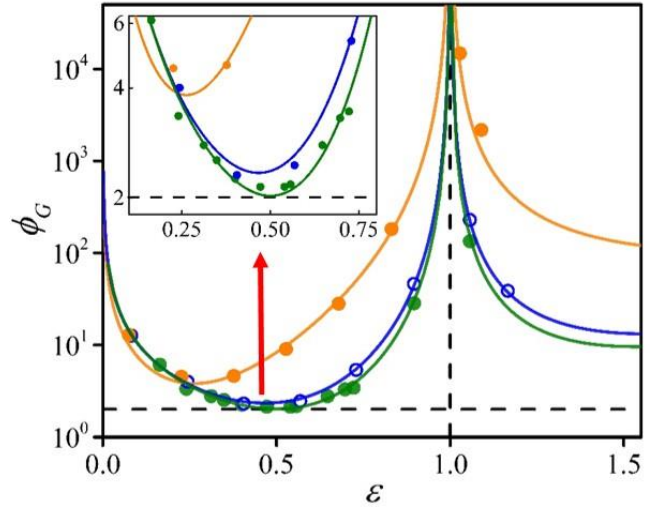

FIG. 2. The measured $\phi_G = Var(\beta W)[\exp(\langle \beta W \rangle + \langle I \rangle) - 1]/\langle \beta W \rangle^2$ as a function of the error level $\varepsilon = \sqrt{N/S}$ for $\tau = 20$ (olive circles), 3 (blue), and 0.5 (orange) ms, for the symmetric feedback. The solid curves are the theoretical model (Eq. (4)). Inset: Expanded view of the main panel showing that $\phi_G$ satisfies the GTUR and achieves the tight bound of $\phi_G^{\min} = 2.03$ for $\tau = 20$ ms and $\varepsilon = 0.5$.

Figure 1(b) shows the experimental test of the TUR prediction, $\phi \geq 2$ (Eq.(3)), for the symmetric feedback, as a function of error level $\varepsilon = \sqrt{N/S}$ at three periods, $\tau = 0.5$, 3, and 20 ms (the characteristic relaxation time is $\tau_R = 3.5$ ms). Faster engines ($\tau \ll \tau_R$), with nonequilibrium initial and final states, always satisfy the TUR, $\phi \geq 2$. Similarly, the TUR is always satisfied for all values of $\tau$ in the $\varepsilon > 1$ region where the average extracted work is negative. However, for $\tau \gtrsim \tau_R$ and for error level in the range $0.28 < \varepsilon < 0.62$, $\phi$ falls below the lower bound set by the TUR (Fig. 1(b) inset). The minimal $\phi$ was found to be about $\approx 1.6$ at $\varepsilon \approx 0.47$ for $\tau \gtrsim 5\tau_R$. The value of $\phi$ diverges near $\varepsilon = 1$ as the average work $\langle \beta W \rangle$ vanishes while the information gain $\langle I \rangle$ remains finite (for error-free measurements, $\varepsilon = 0$, $\phi$ diverges due to the divergence of $\langle I \rangle$). Interestingly, for smaller error level, $\varepsilon < 0.2$, the TUR bound for short period $\tau = 0.5$ ms is lower than that for longer cycle periods (Fig. 1(b) inset).

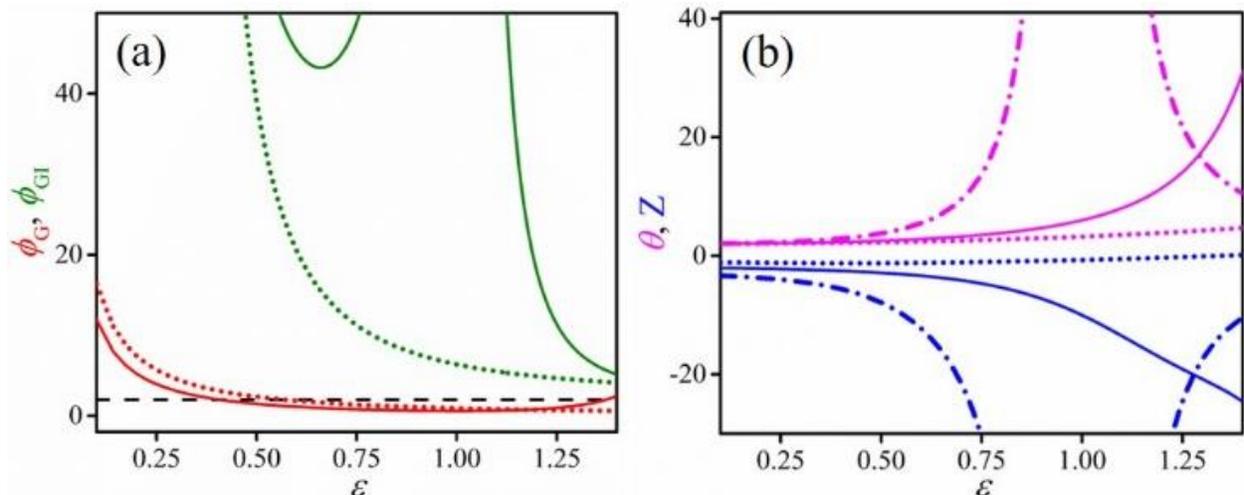

FIG. 3. (a) Plot of $\phi_G$ (Eq. (4)) as a function of error level $\varepsilon = \sqrt{N/S}$ for $\tau = 20$ ms with feedback control of $\lambda = 0.1y$ (red dotted curve) and $0.5y$ (red solid curve). The green curves are the corresponding plots of $\phi_{GI}$ (Eq. (5)). (b) Plot of $\theta \equiv \text{Var}(\beta W)/\langle \beta W \rangle^2$ as of function of $\varepsilon$ for $\lambda = 0.1y$ (magenta dotted curve), $0.5y$ (magenta solid curve), and $y$ (magenta dashed dotted curve). The blue curves are the corresponding plots of $Z$ (Eq. (6)).

In contrast, the TUR is always valid for the asymmetric feedback scheme (Fig. S2 in [34]). In both protocols, the global minimum of the TUR measure $\phi$ is achieved for slow cycles $\tau \gg \tau_R$ where initial and final states are in equilibrium. Note that for the asymmetric feedback scheme, work is extracted only when the measurement outcome is positive ($y \geq 0$). As a result, while the information gained is the same, the average extracted work $\langle \beta W \rangle$ and its fluctuations $\text{Var}\langle \beta W \rangle$ (see [34], Eq. S1) are always less than the symmetric feedback control. This is the reason why TUR is not violated for the asymmetric feedback scheme.

The recently reported *generalized thermodynamic uncertainty relation* (GTUR) sets a softer bound on an observable $X$, $(\text{Var}(X)/\langle X \rangle^2)(e^{\langle \sigma \rangle} - 1) \geq 2$ [22]. The GTUR can be rigorously drived from the strong detailed fluctuation theorem, $P(\sigma)/P(-\sigma) = e^{\sigma}$. It is therefore valid for systems in a periodic steady state for any observable $X$ that is anti-symmetric under time reversal. While this condition may appear to restrict the applicability of the GTUR, we find that it is obeyed in our feedback protocol throughout its phase space (Fig. 2). In our system, the GTUR for the steady state average work current takes the form

$$\phi_G \equiv \frac{\text{Var}(\beta W)}{\langle \beta W \rangle^2}\left[ e^{\langle \beta W \rangle + \langle I \rangle} - 1 \right] \geq 2. \quad (4)$$

The global minimum value of $\phi_G$ (Eq. (4)) is found to be 2.03 at $\varepsilon = 0.5$, for slow engines that fully relax to equilibrium at the end of each cycle (Fig. 2 inset).

The GTUR in Eq. (4) is satisfied for the above feedback protocol of shifting the trap center to the measured outcome, $\lambda = y$. However, we found that $\phi_G$ *falls below the bound* for a general feedback protocol of $\lambda = ay$, for any $0 < a < 1$ (red curves in Fig. 3(a)).

We therefore tested the bound set by another thermodynamic uncertainty relation derived for systems under measurement and feedback control with broken time-reversal symmetry (GITUR1) [32]

$$\phi_{GI} \equiv \frac{\text{Var}(\beta W) + \text{Var}(\beta W)_B}{(\langle \beta W \rangle + \langle \beta W \rangle_B)^2}\left[ e^{(\langle \sigma \rangle + \langle \sigma \rangle_B)/2} - 1 \right] \geq 1, \quad (5)$$

where $\langle \ \rangle_B$ denotes the ensemble average taken over the backward probabilities. Following the backward experiment suggested by the Sagawa and Ueda, where no measurement or feedback is performed, the equilibrium joint probability distributions in forward experiment, $p(x,y)$, and in backward experiment, $p_B(x,y)$, satisfy the generalized detailed fluctuation theorem, $p_B(x,y)/p(x,y) = \exp(-\beta W - I)$ [37]. The corresponding generalized integral fluctuation theorem, $\langle \exp(-\beta W - I) \rangle = 1$, is satisfied by our feedback protocol [30]. Any observables $X$ in the backward experiment can then be calculated as $\langle X \rangle_B = -\int dx dy\, X(x,y) p(x,y) \exp(-\beta W - I)$. For $\lambda = ay$, the information gain at the time of measurement remains the same; however, the work performed on the system during shifting is given by $\beta W = (k/2)[(x-ay)^2 - x^2]$. Unlike the GTUR, the GITUR1 is satisfied for our feedback protocol for all $\lambda = ay$ (green curves in Fig. 3(a)). In particular, the squared relative uncertainty $\theta \equiv \mathrm{Var}(\beta W)/\langle \beta W \rangle^2$ is always bounded by $Z$ ($\theta \geq Z$), where

$$Z \equiv \frac{(1 + \langle \beta W \rangle_B / \langle \beta W \rangle)^2}{e^{(\langle \sigma \rangle + \langle \sigma \rangle_B)/2} - 1} - \frac{\mathrm{Var}(\beta W)_B}{\langle \beta W \rangle^2}, \quad (6)$$

as shown in Fig. 3(b).

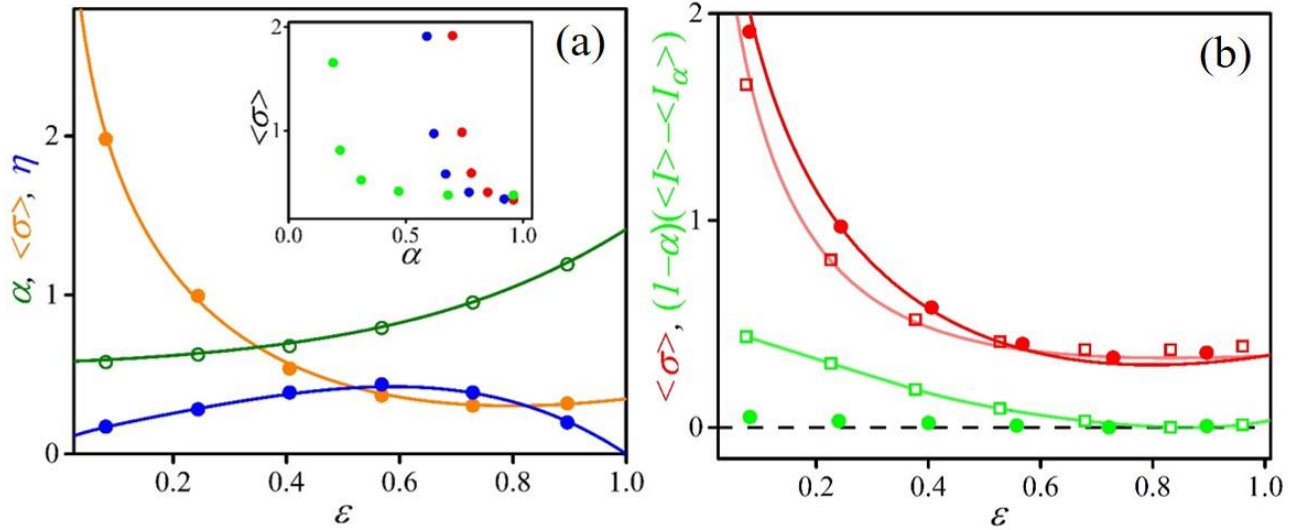

FIG. 4. (a) The normalized work fluctuation $\alpha = std(-\beta W)/std(I)$ (olive), work dissipation $\langle \sigma \rangle = \langle I \rangle + \langle \beta W \rangle$ (orange), and efficiency $\eta = -\langle \beta W \rangle / \langle I \rangle$ (blue) as a function of the error level $\varepsilon = \sqrt{N/S}$ for $\tau = 3$ ms for the symmetric feedback protocol. The solid curves are the theoretical plots. Inset: work dissipation $\langle \sigma \rangle$ as a function of normalized work fluctuation $\alpha$ for $\tau = 20$ (red), 3 (blue), and 0.5 (green) ms. (b) $\langle \sigma \rangle$ (red) and $(1-\alpha)(\langle I \rangle - I_\alpha)$ (green) as a function of error level $\varepsilon$ for $\tau = 3$ (closed circles), and 0.5 (open squares) ms for the symmetric feedback. The red solid curves are the theoretical $\langle \sigma \rangle$. The green solid curve is guide to the eyes. The dashed horizontal line is the second-law bound $\langle \sigma \rangle = 0$.

Note that by choosing the specific backward experiment that includes the measurement and feedback control in the backward path as well [33,38], the GTUR bound in Eq. (4) is also satisfied for the general feedback protocol $\lambda = ay$. The total entropy production for such a protocol should include the entropic cost of the measurement in the backward experiment, $\langle \Sigma \rangle = \langle \Delta S_{sys} \rangle + \langle \Delta S_m \rangle + \langle \Delta S_i \rangle$, where $\langle \Delta S_i \rangle = \langle I \rangle - \langle \ln(p(y|x_\tau)/p(y)) \rangle$ and $x_\tau$ is the particle position at the end of the relaxation (when the next cycle begins for the current feedback protocol) [38]. However, for such backward protocol, the tighter bound on $\theta \equiv \mathrm{Var}(\beta W)/\langle \beta W \rangle^2$ is given by the following generalized thermodynamic relation (GITUR2) [33],

$$\frac{\mathrm{Var}(\beta W)}{\langle \beta W \rangle^2} \geq \mathrm{csch}^2\left[f\left(\frac{\langle \Sigma \rangle}{2}\right)\right], \quad (7)$$

where $f(x)$ is inverse function of $x \tanh(x)$. The bound in Eq. (7) is analogous to the generalized TUR

bound derived from the exchange fluctuation theorem [23]. Eq. (7) is satisfied by our engine, as shown in Fig. S3 in [34]. Interestingly, with this total entropy production $\langle \Sigma \rangle$, the *original* TUR in Eq. (3) is also satisfied for $\lambda = ay$ with $a \geq 0.5$.

*The work fluctuation-dissipation tradeoff and the engine's efficiency.*— The TUR (Eq. (3)) explains a tradeoff between the squared relative uncertainty of an observable current, $\epsilon^2 \equiv \text{Var} X / \langle X \rangle^2$ and the total dissipation, $\langle \sigma \rangle$. It sets a tighter bound to the dissipation ($\langle \sigma \rangle \geq 2/\epsilon^2$) than that set by the second law of thermodynamics ($\langle \sigma \rangle \geq 0$). However, as shown above, information engines often fall below the TUR bounds. Very recently, Funo and Ueda reported a general tradeoff relation (denoted here IDR, information distance relation) between the fluctuation of the extracted work and the dissipation in an information engine [31]:

$$\langle \sigma \rangle \geq (1-\alpha)(\langle I \rangle - \langle I_\alpha \rangle), \qquad (8)$$

where $\alpha = std(\beta W) / std(I)$ is the work fluctuation normalized by the information fluctuation, and $\langle I_\alpha \rangle = \left( \ln \sum_{x,y} p(x,y)^\alpha [p(x)p(y)]^{1-\alpha} \right) / (\alpha - 1)$ is the Renyi generalized mutual information. Fig. 4(a) shows the tradeoff between the normalized work fluctuation $\alpha$ and the dissipation $\langle \sigma \rangle$ in the positive average work extracting region of the engine ($\varepsilon < 1$), for $\tau = 3$ ms. Similar tradeoff behavior is observed for all $\tau$, particularly for $\alpha \leq 1$ which correspond to $\varepsilon \leq 0.74$ (see Fig. 4(a) inset).

Finding the protocol that simultaneously minimizes the dissipation and the uncertainty in the extracted work is crucial for the design of efficient engines. The dissipation is minimal for slower engines at finite error level $\varepsilon \approx 0.78$. The information utilization efficiency, $\eta \equiv -\langle \beta W \rangle / \langle I \rangle$ of this engine is found to be maximal for slower engine at $\varepsilon \approx 0.6$, close to the minimal dissipation point (Fig. 4(a)). Note that the original TUR is violated near this error level of maximal efficiency. We also demonstrated that the IDR trade off in Eq. (8) is always satisfied for our protocol as shown in Fig. 4(b). The tighter bound can be achieved by a protocol optimized for maximal work extraction, such as one described in [39]. The optimal protocol combines an instantaneous shift of the trap center to a new position $y \cdot S/(S+N)$ and a simultaneous stiffening of the trap $k \to k' = (1 + S/N) \cdot k$, followed by an adiabatic softening back to the original spring constant, $k' \to k$. With this protocol all the available information is utilized as work $\langle -\beta W \rangle = \langle I \rangle$ and the dissipation vanishes, $\langle \sigma \rangle = 0$. The work fluctuation remains unity, $\alpha = 1$, irrespective of error size, thus achieving the sharp IDR (an equality in Eq. (8)).

*Conclusion.*— The original TUR provides a fundamental lower bound on the fluctuation-dissipation tradeoff of non-equilibrium processes. This bound constrains the efficiency of the system. We show that, in the certain range of the parameters, the information engine violates the original TUR and satisfies the softer GTUR bounds only for $\lambda = y$, while for $\lambda = ay$, the GTUR violated. We show that the engine always satisfies the GITUR and the IDR bounds. The GITUR requires the design of backward experiments in which the thermodynamic observables and their fluctuations are measured along the backward trajectory. However, realizing the backward experiments for a cyclic information engine operating in a non-equilibrium steady state is often challenging. Nevertheless, the IDR bound that links the work fluctuation, mutual information and Renyi entropy may serve as an alternative uncertainty relation for the work fluctuation-dissipation tradeoff. The role of fluctuation and dissipation in shaping the efficiency of the engine was also studied, and we found that the tradeoff gives rise to a peak efficiency when the dissipation is minimal or when the fluctuation and dissipation are of similar magnitude. The present work may inspire further studies on the connections between uncertainty relations and the efficiency bounds in feedback systems. Finally, we note that a recently-reported generalized TUR, for systems with arbitrary initial states, suggests original TUR is violated in our information engine because the average work current per cycle decreases with time due to relaxation, as a result, the instantaneous current at the end of the relaxation is always smaller than the total current per cycle [40].

This work was supported by the taxpayers of South Korea through the Institute for Basic Science, project code IBS-R020-D1.

†tsvitlusty@gmail.com

*hyuk.k.pak@gmail.com

**Supplementary Material**

**Figures**

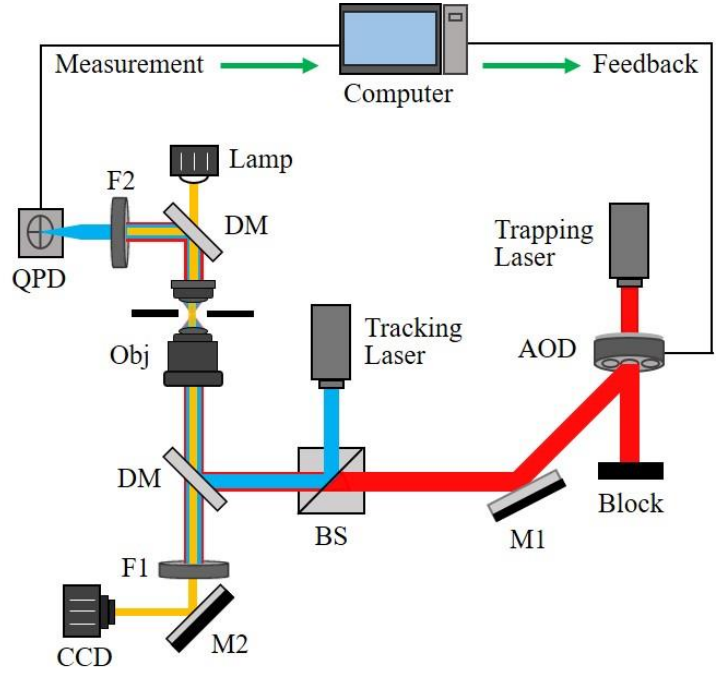

FIG. S1. Basic schematics of the optical tweezers set up with real time feedback control. The detailed schematics can be found in Ref. [3].

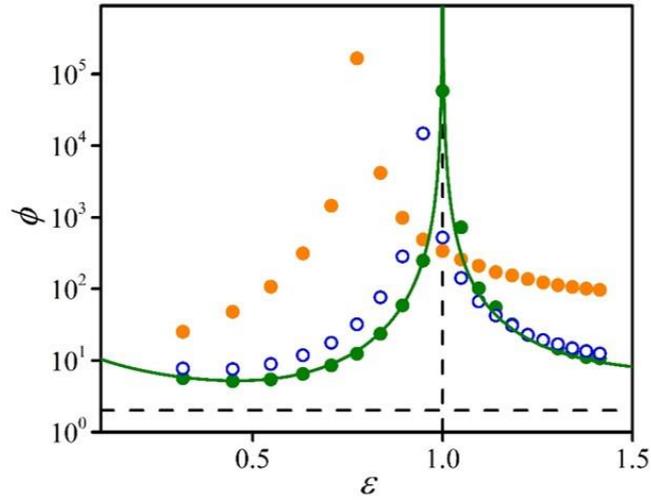

FIG. S2. (a) Test of TUR for asymmetric feedback scheme obtained from simulation for $\tau = 20$, (olive circles), 3 (blue), and 0.5 (orange) ms. The solid curve is obtained from analytical results (Eq. (S3)).

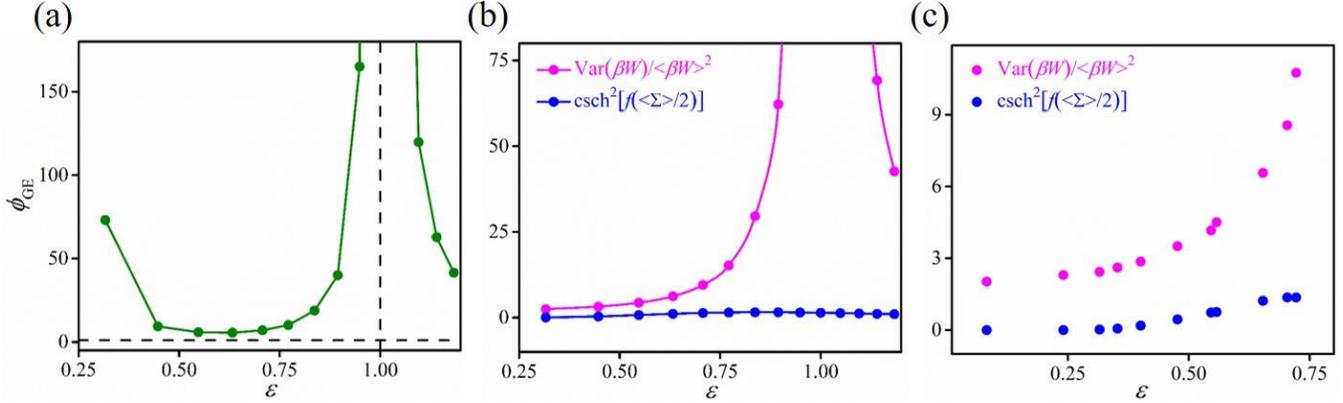

FIG. S3. Test of GITUR2 in Eq. (7) obtained from simulation. (a) Plot of $\phi_{GE} \equiv (\mathrm{Var}(\beta W)/\langle \beta W \rangle^2)(1/\mathrm{csch}^2[f(\langle \Sigma \rangle/2)])$ as a function of the error level $\varepsilon = \sqrt{N/S}$ for $\tau = 20$ ms and $\lambda = y$. Note the GITUR2 in Eq. (7) is violated for $\langle \sigma \rangle = \langle \beta W \rangle + \langle I \rangle$. (b) Plot of the square relative uncertainty $\theta = \mathrm{Var}(\beta W)/\langle \beta W \rangle^2$ (magenta), and the lower bounds of the GITUR2 ($\mathrm{csch}^2[f(\langle \Sigma \rangle/2)]$, blue) as a function of the error level $\varepsilon$ for the data in panel (a). The solid curves are guide to the eye. (c) Experimental verification of Eq. (7) for $\tau = 20$ ms and $\lambda = y$.

## Materials and Methods

**Experimental setup.** The schematic of the optical tweezers set up with real time feedback control is similar to our recently published works [1-3] (Fig. S1). A laser with 1064 nm wavelength is used for trapping the particle. The laser is fed to an acoustic optical deflector (AOD) via an isolator and a beam expander. The AOD is controlled via an analog voltage controlled radio-frequency (RF) synthesizer driver. The AOD is properly mounted at the back focal plane of the objective lens so that $k$ is essentially constant while shifting the potential center. A second laser with 980 nm wavelength is used for tracking the particle position. A quadrant photo diode (QPD) is used to detect the particle position. The electrical signal from QPD is amplified and sampled at every $\tau$ with a field-programmable gate array (FPGA) data acquisition card. The sample cell consist of highly dilute solution of 2.0 μm diameter polystyrene particles suspended in deionized water. All experiments were carried out at $293 \pm 0.1$ K. The parameters of the trap were calibrated by fitting the probability distribution of the particle position in thermal equilibrium without a feedback process to the Boltzmann distribution $P(x) = (2\pi\sigma^2)^{-1/2} \exp(-x^2/2\sigma^2)$. The trap stiffness is then obtained by using the relation $k = k_B T/\sigma^2 \approx 5.40$ pNum$^{-1}$.

## TUR bounds for asymmetric feedback

The $i$-th engine cycle under a*symmetric* feedback control operates by measuring the *true* particle position $x_i$ with respect to the potential center $\lambda_{i-1}$, as $y_i$. The trap center is then shifted instantaneously to $y_i$ only if $y_i \geq \lambda_{i-1}$, and otherwise remains at $\lambda_{i-1}$ until the next cycle. The average work performed on the particle per cycle $\langle \beta W \rangle$ and its standrad deviation $\mathrm{std}(\beta W)$ for large cycle period ($\tau \geq 5\tau_R$) are given by

$$\langle \beta W \rangle = \frac{1}{2}\beta k \int_{-\infty}^{\infty} dx \int_0^{\infty} dy\, p(x|y)p(y)\left[(x-y)^2 - x^2\right]$$
$$= -\frac{1}{4}\left(\frac{S-N}{S}\right),$$

$$\text{std}(\beta W) = \sqrt{\frac{5(N^2 + S^2) - 2NS}{16S^2}} \tag{S1}$$

Similarly, the average mutual information gain per cycle $\langle I \rangle$ is

$$\begin{aligned}\langle I \rangle &= \int_{-\infty}^{\infty} dx \int_{-\infty}^{\infty} dy\, p(x|y) p(y) \ln \frac{p(x|y)}{p(x)} \\ &= \frac{1}{2}\ln\left(1 + \frac{S}{N}\right),\end{aligned} \tag{S2}$$

The resulting TUR [4-7] for the average power per cycle, $P \equiv \langle \beta W \rangle / \tau$ is

$$\phi \equiv \frac{\text{Var}(\beta W)}{\langle \beta W \rangle}\left(1 + \frac{\langle I \rangle}{\langle \beta W \rangle}\right) \geq 2. \tag{S3}$$

We found that the TUR in Eq. (S3) is always valid for the asymmetric feedback control (Fig. S1).